\documentclass[sigconf, authorversion,nonacm]{acmart}

\AtBeginDocument{%
  }

\newcommand{\indsize}{\scriptsize}
\newcommand{\colind}[2]{\displaystyle\smash{\mathop{#1}^{\raisebox{.5\normalbaselineskip}{\indsize #2}}}}
\newcommand{\rowind}[1]{\mbox{\indsize #1}}
\usepackage{subfig}
\usepackage{graphicx}

\begin{document}

\title{MGS: Markov Greedy Sums for Accurate Low-Bitwidth Floating-Point Accumulation
}

\author{Vikas Natesh}
\email{vnatesh@g.harvard.edu}
\orcid{}
\affiliation{%
  \institution{Harvard University}
  \streetaddress{P.O. Box 1212}
  \city{Cambridge}
  \state{MA}
  \country{USA}
}

\author{H.T. Kung}
\email{kung@harvard.edu}
\orcid{}
\affiliation{%
  \institution{Harvard University}
  \streetaddress{P.O. Box 1212}
  \city{Cambridge}
  \state{MA}
  \country{USA}
}

\author{David Kong}
\email{dkong@g.harvard.edu }
\orcid{}
\affiliation{%
  \institution{Harvard University}
  \streetaddress{P.O. Box 1212}
  \city{Cambridge}
  \state{MA}
  \country{USA}
}



\begin{abstract}
We offer a novel approach, MGS (Markov Greedy Sums), to improve the accuracy of low-bitwidth floating-point dot products in neural network computations. 
In conventional 32-bit floating-point summation, adding values with different exponents may lead to loss of precision in the mantissa of the smaller term, which is right-shifted to align with the larger term's exponent. Such shifting (a.k.a. 'swamping') is a significant source of numerical errors in accumulation when implementing low-bitwidth dot products (e.g., 8-bit floating point) as the mantissa has a small number of bits. We avoid most swamping errors by arranging the terms in dot product summation based on their exponents and summing the mantissas without overflowing the low-bitwidth accumulator. We design, analyze, and implement the algorithm to minimize 8-bit floating point error at inference time for several neural networks. In contrast to traditional sequential summation, our method has significantly lowered numerical errors, achieving classification accuracy on par with high-precision floating-point baselines for multiple image classification tasks.
Our dMAC hardware units can reduce power consumption by up to 34.1\% relative to conventional MAC units.

\end{abstract}

\maketitle

\section{Introduction}

Quantization has become a ubiquitous optimization for compressing deep neural networks (DNNs) on both low-power edge devices \cite{jorge,siracusa,mcunet,mema} as well as large-scale training and inference systems made up of many GPUs \cite{msft}.
Low-power devices for tinyML typically have small local memories \cite{m4} and often lack support for efficient floating-point computation \cite{gap8,siracusa}.
Hence, integer quantization is, by default, necessary on such systems, and most tinyML models are quantized to 8 bits or less.
Meanwhile, large generative AI workloads push GPU-based training and inference clusters to the limits of available memory, bandwidth, and computation power.
To this end, low-bitwidth formats such as brain float-16 (bfloat16) \cite{bfloat}, block floating point (BFP) \cite{bfp}, and 8-bit floating-point (FP8) have been implemented in various hardware \cite{h100, gaudi}.
Such formats have been successful in reducing memory footprint, memory accesses, computation time, and power consumption \cite{h100, gaudi, fp8}

\begin{figure}[h!]
\centering
\includegraphics[width=\linewidth]{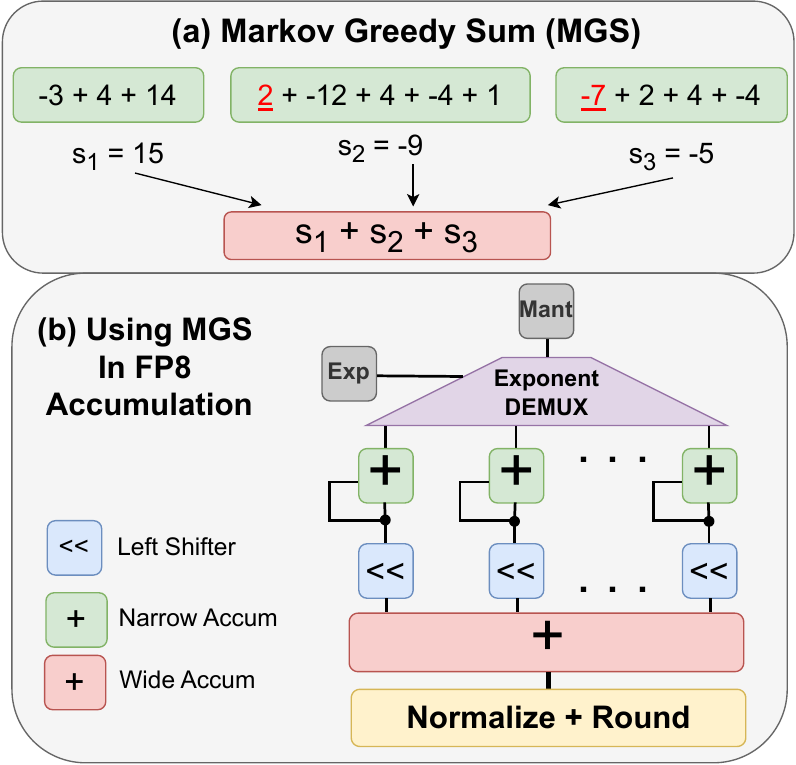}

\caption{
(a) Example of Markov Greedy Sums (MGS).  In (a), we sum 12 integers into a narrow accumulator (green box) until the sum $s_i$ overflows the range [-15, 15].
Then, we accumulate $s_i$ into a wider accumulator (red box).
The underlined red values are those that would have caused an overflow of the narrow accumulator, noting that 15 + 2 > 15 and -9-7 < -15.
In (b), we accumulate FP8 values by separating them by their exponents into 16 groups, summing the mantissas into separate narrow accumulators indexed by the exponent, and using a wide accumulator upon overflow.
MGS amortizes the cost of aligning (shifting) FP8 mantissas over many sums. 
}
\label{fig:overview}
\end{figure}

When performing quantized matrix multiplications, dot products are typically accumulated into wider registers.
For instance, partial products in FP8 may be accumulated in FP16 or FP32 to ensure numerical accuracy.
Reducing accumulator bitwidth can reduce bandwidth and energy usage while increasing inference throughput \cite{a2q, a2q+, wrapnet,henk}.
However, if the partial product sum overflows the accumulator, its value may be clipped to a finite range.
This introduces numerical errors into the final matrix result that degrade model accuracy and limit how much one can reduce the accumulator bitwidth.
In addition, there is the swamping problem \cite{higham} that causes loss of precision due to mantissa shift when floating point numbers need to align their exponents before summation. 

Prior works have attempted to reduce overflow in narrow accumulators by regularizing the loss function  \cite{wrapnet} or by controlling weight magnitude during training \cite{a2q,a2q+,henk}. 
While these approaches succeed in reducing overflows, they impose restrictive constraints on weights during training that may reduce model accuracy \cite{a2q, a2q+, wrapnet}.
Moreover, existing networks require expensive re-training to satisfy such constraints.
Other works reorder dot product summations to avoid the majority of overflows when using narrow accumulators \cite{ags}.
However, reordering requires additional sort/permute operations as well as memory for temporary storage and is difficult to optimize on existing hardware.

We propose \textbf{M}arkov \textbf{G}reedy \textbf{S}ummation (MGS), a novel approach to enable low-precision accumulation in neural network dot products without the need for retraining or summation reordering.
We analyze overflows during neural network inference and model the value of the partial sum in dot products as a Markov process to derive the expected dot product length without overflow.
Our key insight is that based on the statistical properties of weight and activation distributions, we can sum many partial products in reduced precision before overflow occurs. 
MGS is greedy in the sense that it uses a narrow accumulator to accumulate as many values as possible while falling back on a wider accumulator when the rare overflow occurs.
Leveraging this insight, we design dual-multiply-accumulate (dMAC) hardware units that use narrow accumulators for the majority of sums.
Our method uses a narrower average accumulator bitwidth compared to prior works when performing DNN computations.
Our dMAC units consume up to 34.1\% less energy than conventional integer and floating-point MACs that use wide accumulators for all summations.
Figure \ref{fig:overview} provides an overview of MGS applied to integer and FP8 summation.
The novel contributions of this paper are:
\begin{itemize}
    \item Analysis of dot product overflows in integer and floating-point quantized neural networks (Section \ref{sec:analysis} ).
    \item Dual-MAC hardware architecture (dMAC) and algorithm for avoiding overflows when using narrow accumulators in integer and FP8 dot products (Sections and \ref{sec:mgs} and \ref{sec:dmac}).
    \item Evaluation of the resulting methods regarding model accuracy and accumulator compression for classification tasks. We emulate our integer and FP8 dMAC units on both CPU and GPU hardware and evaluate task accuracy and accumulator compression for multiple DNNs (Section  \ref{sec:eval}).
    \item Energy consumption and area evaluation of dMAC units compared to conventional integer and floating point MACs when implemented in a 7nm process node. (Section \ref{sec:eval})
\end{itemize}

\section{Background}
\label{sec:background}

We present background on both integer and floating-point quantization of DNNs and some prior work on avoiding overflow during DNN execution.

\subsection{Integer Quantization}
We consider the uniform quantization of weight and activation matrices per-tensor from FP32 to $b$-bit signed values \cite{jacob}.
The floating-point values in a matrix $M$ have a range $R = max(X) - min(X)$.
To map values in $M$ to integers in $[0, 2^b - 1]$, we partition $R$ into $2^b - 1$ uniform intervals of length $s_x = \frac{R}{2^b -1}$, also called the scale factor.
For example, we can map a FP32 activation $x$ to a value $x^q$ in $[0, 2^b - 1]$ using the equation $x^q = round(\frac{x^f}{s_x})$.
If the range is asymmetric around zero, we shift $x^q$ by an offset $o_x = -2^{b-1} - round(\frac{min(X)}{s_x})$ into the range $[-2^{b-1}, 2^{b-1} - 1]$, guaranteeing that the FP32 value for 0 maps to an integer.
We can obtain the approximate FP32 representation of a quantized activation $x^q$ by reversing the effect of the scale and offset via the equation $x^* = s_x(x^q - o_x)$.
Quantized dot products are then performed using the FP32 approximations.
$$s_z(z - o_z) = \sum_{i=1}^{K} s_w(w_i^q - o_w) s_x(x_i^q - o_x)$$
 where $s_w$, $o_w$, $s_z$, and $o_z$ represent the quantization parameters of weights $w$ and output activations $z$.
Floating point scale factors are factored out and normalized to an integer representation, while weights are typically symmetric around zero with $o_w = 0$ \cite{jacob,tensorflow,termquant,pytorch}.
As a result, several terms under the summation disappear, and the majority of computation arises from the integer dot product $z = \sum_{i=1}^{K} w_i^q x_i^q$.

When FP32 weights and activations are quantized to low-precision (e.g., 8-bit), the computation cost of multiplications $w_i^q x_i^q$ decreases significantly. 
However, the compute bottleneck transitions to the $K$ dot product summations, as these accumulations are typically executed in higher precision, such as 32-bit, to avoid overflow of the accumulator.
For example, assume we accumulate using a $p$-bit register where each partial product $w_i^q x_i^q$ is $2b$-bits and $p > 2b$. 
This leaves $p - 2b$ bits leftover for precision during accumulation. 
Hence, the dot product overflows when $K\geq2^{p - 2b}$.
However, if we use a narrow accumulator $p = 2b$, overflow may occur during any of the $K$ partial sums, leading to inaccurate dot product and poor model accuracy.

Previous works enable the use of narrow accumulators in DNN computations by retraining the network to reduce partial sum magnitude \cite{wrapnet, a2q, a2q+} or algorithmically avoiding most overflows \cite{ags}.
In practice, ML frameworks for quantized DNNs avoid overflow by either using high-precision accumulators (e.g., 32-64 bits) or clipping partial results into a finite range (saturation arithmetic) as they are accumulated \cite{neon, cmsis1, pulpnn}.
Clipping is cheap to implement in hardware or software, allowing for a modest reduction in accumulator precision, e.g., from 32 to 16 bits.
However, for narrower bitwidths (< 16), clipping severely degrades numerical accuracy and task performance \cite{a2q, ags}.

\begin{figure}[h!]
\centering
\includegraphics[width=\linewidth]{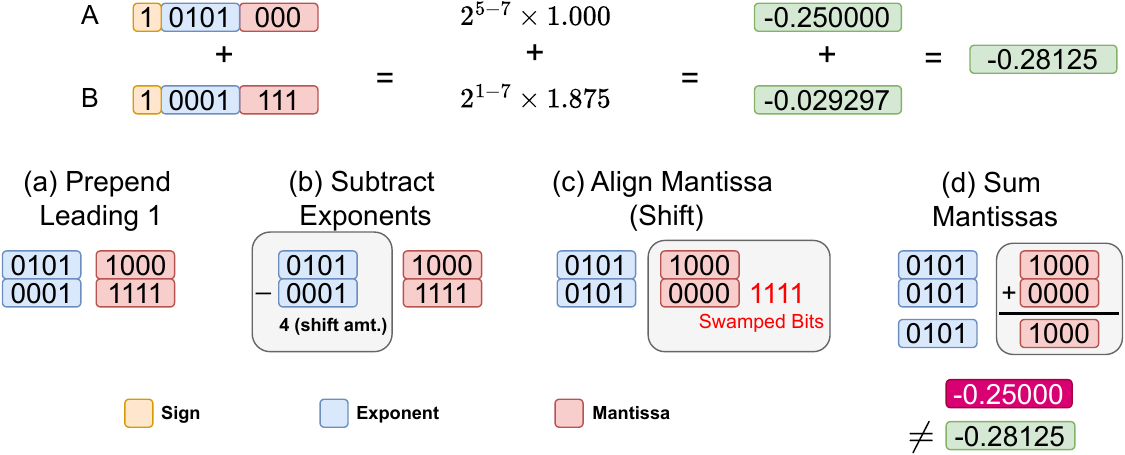}

\caption{
An example of mantissa bit swamping when adding two E4M3 values with different exponents, $A =-0.25$ and $B=-0.029297$, while using a narrow 4-bit accumulator.
The exponent bias in E4M3 is 7.
A's exponent of 5 is larger than B's exponent 1 (b), causing B's mantissa to be shifted left by 5-1=4 bits (c).
Since the entire mantissa shifts out, B is treated as zero, and the final result is 0.25, differing from the closest FP8 result of -0.28125 (d).
}
\label{fig:swamping}
\end{figure}

\subsection{Floating-Point Quantization}
FP32 DNN weights and activations may be quantized to lower-precision floating-point formats such as bfloat16, BFP, FP16, or FP8.
In particular, FP8 formats for both inference and training have been developed and implemented on several commercial AI accelerators, such as Nvidia H100 GPUs and Intel Gaudi2 \cite{h100, gaudi, fp8}. 
Such formats are now widely used and can achieve baseline FP32 performance on large AI workloads, such as LLMs \cite{msft, deepseek}.

FP8 summation involves several steps as shown in Figure \ref{fig:swamping}.
Consider the E4M3 format with one sign bit, four exponent bits, and three mantissa bits \cite{fp8}.
When adding two FP8 values with different exponents, the lower order bits of the smaller value are shifted out (`swamped') due to right-shifting to align exponents with the larger value.
This leads to a loss of precision in the final sum.
In contrast to floating-point formats with wide mantissas, narrow formats such as E4M3 suffer from a significant loss in numerical accuracy due to swamping.
Commercial hardware such as the H100 avoids swamping by accumulating FP8 partial sums in a wider precision such as FP16 or FP32 \cite{h100}.

There are several classical algorithms for reducing swamping error in floating point summation, including pairwise summation \cite{higham} and Kahan summation \cite{kahan}.
Although Kahan summation has higher accuracy, it requires several extra floating point operations to maintain the compensated error term.
Meanwhile, pairwise summation is efficient to implement but suffers from large error in narrow floating-point formats.

\begin{figure}[h!]
\centering
\includegraphics[width=\linewidth]{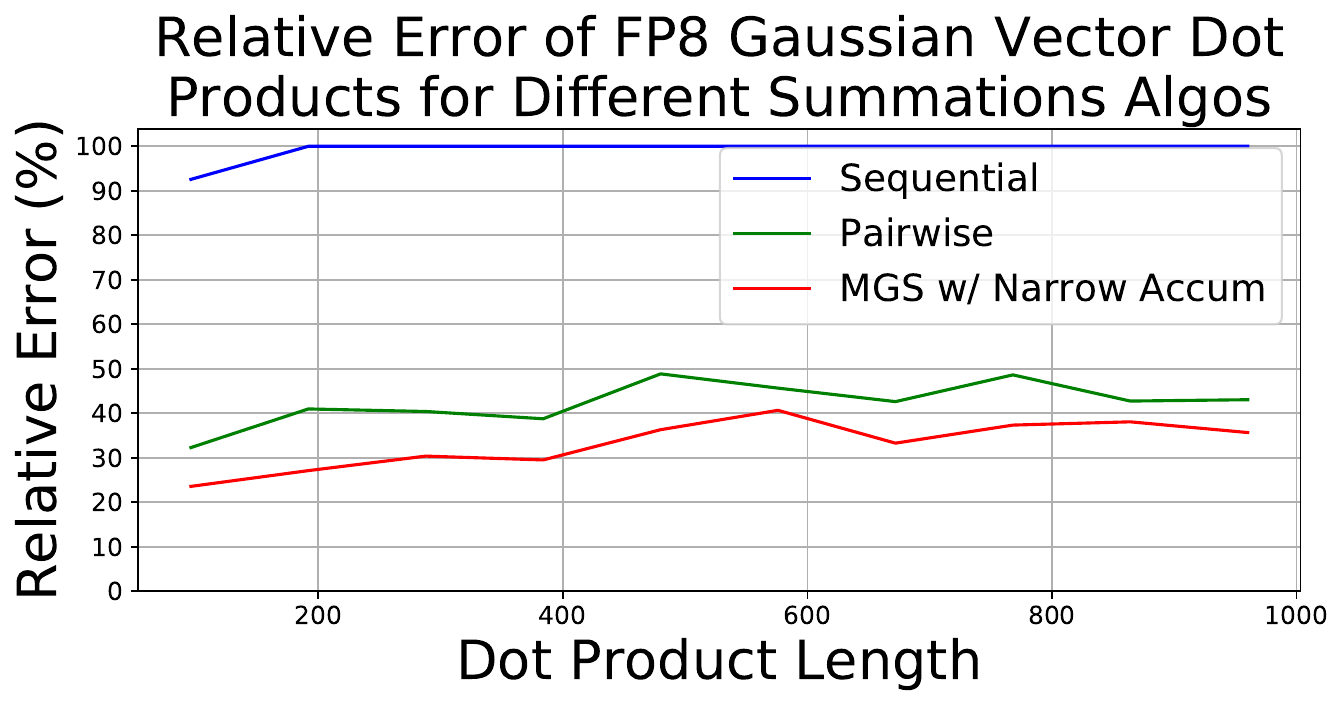}

\caption{
\% Error, relative to FP32 precision, of Gaussian vector dot products performed in FP8 precision.
We execute each algorithm using solely a narrow accumulator and clip partial sums upon overflow.
All algorithms exhibit significant errors due to the swamping of lower order bits when using reduced-precision accumulators.
MGS has lower error than pairwise summation by separating partial product mantissas by exponent and accumulating them in separate narrow accumulators.
This means that dot product errors result only from clipping overflows.
However, the $\approx 35\%$ error of MGS, when restricted to a narrow accumulator, is unacceptable for DNN applications.
}
\label{fig:error}
\end{figure}

Figure \ref{fig:error} illustrates the need for high-precision accumulation of FP8 dot products.
Using several summation algorithms, we perform dot products between two Gaussian vectors in FP8 precision (4-bit mantissa accumulator) and plot the numerical error relative to the baseline FP32 accumulation (24-bit mantissa accumulator).
Sequential summation loses all accuracy after only 200 sums.
Pairwise summation is significantly more accurate than sequential summation but still exhibits up to 50\% error for longer dot products.
In Section \ref{sec:fp8}, we discuss how to accumulate FP8 mantissas in low-precision for a majority of sums while attaining numerical accuracy on-par with FP32 accumulation.

\section{Analysis of Dot-Product Overflows}
\label{sec:analysis}
We begin by providing an analytical framework for reasoning about overflows. 
We define two types of integer overflow and discuss multiple algorithms for avoiding them. 
\begin{definition}[Transient Overflow]
Overflow that may occur at any point during the sequential summation of $k$ integers $X = \{x_1, x_2, ..., x_k\}$  when using a $b$-bit accumulator.
\end{definition}

\begin{definition}[Persistent Overflow]
Overflow that occurs when the final sum $y = \sum_{i=1}^{k} x_i$ overflows a $b$-bit accumulator.
\end{definition}

Note that transient overflows may occur even when there is no persistent overflow. We aim to minimize these transient overflows.

\subsection{Avoiding All Overflows}

Several prior works aim to avoid both persistent and transient overflows entirely by retraining the neural network such that partial sums are always within the accumulator bounds.
A2Q \cite{a2q} and A2Q+ \cite{a2q+} eliminate the possibility of both transient and persistent overflows by constraining the weight vector's L1-norm during quantization-aware training (QAT). They first bound the dot product result :
\[  |\sum_{i=1}^{K} w_i x_i| \leq \sum_{i=1}^{K} |w_i| |x_i| \leq 2^{p-1} - 1\]
In the worst case, all activations are maximal $|x_i| = 2^{b-1}$ and the weight L1-norm may be bounded such that:
\[ \sum_{i=1}^{k} |w_i| = \| \mathbf{w}\|_1 \leq \frac{2^{p-1} - 1}{2^{b-1}}\]
This bound acts as an L1-regularizer and pulls most weight values toward zero, ensuring that partial sums never grow beyond $p$ bits.
L1 regularization promotes unstructured sparsity in the weight matrices, reducing the model size and enabling acceleration by skipping zero computations.
However, network sparsification may reduce model accuracy \cite{malach} 
Meanwhile, retraining a pre-trained DNN to satisfy accumulator constraints may alter properties of the pre-trained model, such as algorithmic fairness guarantees \cite{fair}.
We find that enforcing strict bounds on weight magnitude is not necessary for using narrow accumulators.

\subsection{Avoiding Transient Overflows}

Persistent overflow is a true overflow where the final result is simply too large for the accumulator.
Transient overflows are `temporary' and arise when a partial sum overflows but where the final sum may not actually overflow the accumulator.
Hence, in the absence of persistent overflow, we should be able to eliminate transient overflows by reordering the summation.
\begin{theorem}
Let $X = \{x_1, x_2, ..., x_k\}$ be a list of $k$ signed integers, where each $x_i$ is represented using $n$ bits. Let $y = \sum_{i=1}^{k} x_i$ be the sum of all elements in $X$, representable using $m \geq n + 1$ bits without persistent overflow (i.e., $-2^{m-1} \leq y \leq 2^{m-1} - 1$). Then, there exists an ordering of summation for $X$ that avoids transient overflow when using an $m$-bit accumulator.\end{theorem}

\begin{proof}
Suppose $k=2$. The list $X = \{x_1, x_2\}$ contains $n$-bit numbers, and its sum $x_1 + x_2$ (or $x_2 + x_1$) can require at most $n+1$ bits. Since $m\geq n+1$, the sum does not overflow $m$-bits, and the theorem holds for $k=2$.


Let $l \geq 2$ and assume the theorem holds $\forall{k} \leq l$, i.e., there exists an ordering of $X = \{x_1, x_2, ..., x_l\}$ such that the sum of elements of $X$ w.r.t. said ordering avoids transient overflow (inductive hypothesis). 
Denote this ordering by the index set $\alpha_l$ and the so-ordered list by $X_{\alpha_l}$.
Suppose that $k = l + 1$ and $X = X_{\alpha_l} \cup x_k$. Then,
\[  y = \sum_{i=1}^{k} x_i = x_k + \sum_{i \in \alpha_l} x_i \]

By the inductive hypothesis, the second term in the sum, denoted by $\hat{y} = \sum_{i \in \alpha_l} x_i$,  avoids transient overflow. Since $\hat{y}$ is an $m$-bit signed integer, and $x_k$ is an $n$-bit signed integer with $n \leq m - 1$, the sum $y = x_k + \hat{y}$ is represented by at most $m$ bits. Therefore, a feasible ordering to avoid transient overflow is $\alpha_k = \{\alpha_l, k\}$. Thus, by induction, our theorem must hold for any $k \geq 2$.
\end{proof}

The proof shows how to construct a summation sequence without transient overflow by building the `right' permutation sequence at each step.
One example of such an algorithm is first to sort the $k$ values, divide them into a list of negative values and a list of positive values, and repeatedly form the sum of the largest positive and most negative values.
We can then take the resulting list, with length at least $k/2$, and apply the algorithm recursively until a single pair of values remains.
This method is guaranteed to avoid transient overflow while using the narrowest possible accumulator as the running sum increases monotonically.
Performing summations in a sorted order is also beneficial for retaining floating point accuracy since adding pairs of values of similar magnitude reduces the number of bits swamped in the smaller value \cite{dix}.
However, sorting before adding becomes expensive in DNN applications where dot product lengths may exceed 4096.

AGS is a recent method to avoid transient overflow by reordering in integer-quantized DNNs \cite{ags}. 
AGS first splits the sequence by sign into a positive list and negative list, then alternates summing values from either the negative or positive list depending on whether the accumulator overflows its maximum or minimum value, respectively.
This allows AGS to avoid transient overflows while also avoiding sorting and using only an extra bit for overflow detection.
However, AGS may require additional registers or memory to buffer partial products.
For example, once an overflow is detected, AGS may need to buffer several positive values while waiting for a negative partial product to arrive.
The extra memory requirements may overwhelm the benefits of using a narrow accumulator, challenging AGS hardware implementation.

\section{Markov Greedy Summation}
\label{sec:mgs}
In this section, we detail how our proposed MGS avoids all overflows while using narrow accumulators for the majority of summations.
We show that since weights and activations are normally distributed or half-normally distributed, the chance of overflow during summation is actually low.
We then derive the expected number of summations before overflow by modeling the running sum as a random walk.\\

\begin{figure}[h!]
\centering

    \subfloat{\label{fig:contour1}{\includegraphics[width=0.48\linewidth]{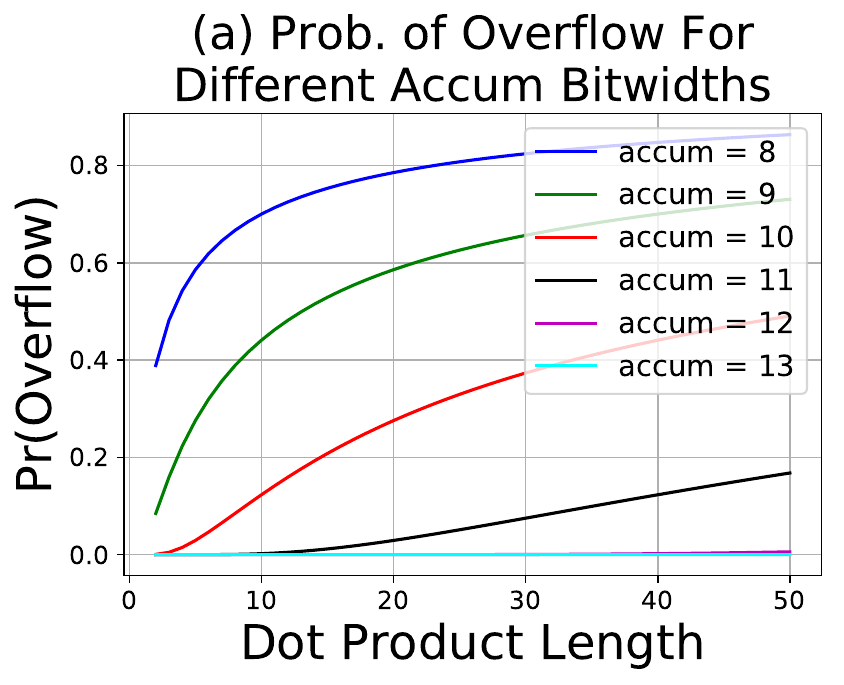}}}
    \subfloat{\label{fig:contour2}{\includegraphics[width=0.5\linewidth]{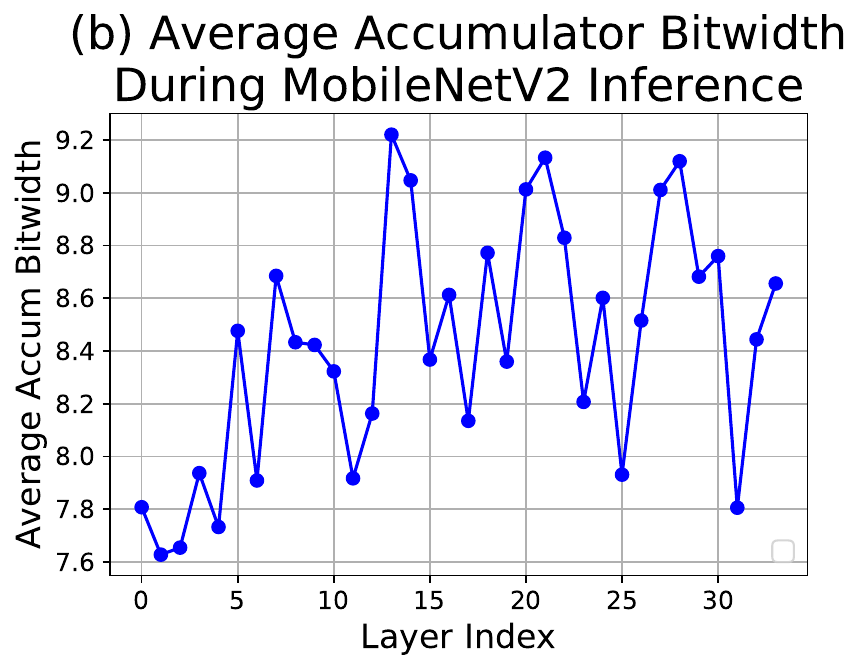}}}

\caption{
(a) We estimate the probability of overflow based on the model described in Section \ref{sec:prob}, when performing dot product at different accumulator bitwidths.
5-bit Gaussian weights in the range[-15,15] are multiplied with 7-bit Gaussian activations in [-63,63] to yield partial products $Z \approx N(0, k * \sigma_w \sigma_x)$.
We set $\sigma$ of weights and data such that the extreme values lie 3 $\sigma$'s away from the mean 0, i.e., $\sigma_w = 15/3 = 5$ and $\sigma_x = 63/3 = 21$.
The figure shows that despite 7+5=12-bit partial products, we can use accumulators with < 12 bits for most sums before overflow.
For example, there is only a $\approx$ 12\% chance of overflow when summing 10 elements in a narrow 10-bit accumulator.
In (b), we plot the average accumulator bitwidth when running MobileNetv2 inference with 5-bit weights and 7-bit activations.
Although one would expect that at least 5+7=12 bits are required to prevent overflow, the average accumulator bitwidth required varies between 7 and 10 bits.
} 
\label{fig:clt}
\end{figure}

\subsection{Estimating the Probability of Integer Overflow}
\label{sec:prob}
We consider $b$-bit quantized neural network dot products $Z = \sum_{i=1}^{k} w_i x_i$ with weights and activations in the range $[-2^{b-1}, 2^{b-1}]$.
Weight and input activation vectors $w$ and $x$ are truncated, zero-centered i.i.d normal distributions $N(\mu_w =0, \sigma_w)$ and $N(\mu_x = 0, \sigma_x)$, respectively.
Input activations may also be half-normal distributions due to ReLU operations in the previous layer.
The partial products $p_i = w_i  x_i$ are i.i.d product-normal distributions with $\mu_p = 0$ and $\sigma_p^2 = (\sigma_w^2 + \mu_w^2) (\sigma_x^2 + \mu_x^2) - \mu_w^2 \mu_x^2 = \sigma_w^2 \sigma_x^2$.
The summing of partial products can be represented by the random variable $Z = \sum_{i=1}^{k} p_i$.
By the central limit theorem (CLT), for large enough $k$, $Z \approx N(0, k * \sigma_w \sigma_x)$. This enables us to approximate the probability of overflow given a particular dot product length $k$ and accumulator bitwidth $a$.
$$Pr(|Z| > 2^{a-1}) \approx 2\Phi \left[ {\frac{-2^{a-1}} {\sigma_w \sigma_x \sqrt{k}} } \right]$$
 where $\Phi$ is the CDF of the standard normal distribution.
Figure \ref{fig:clt}a displays the probability of overflow for different vector lengths and accumulator bitwidths when performing dot product with 5-bit weights and 7-bit activations.
The figure shows that for relatively long dot products, such as 10 or 15 elements, the chance of overflow is relatively low, even for narrow accumulators.
In Figure \ref{fig:clt}b, we empirically observe that the average accumulator bitwidth is small across DNN layers, suggesting that wide accumulators may not be needed for a majority of sums.

\subsection{Computing the Expected Number of Overflows}
The approximation above provides a loose bound showing that overflow is relatively rare, even with narrow accumulators.
We can derive the expected number of summations before overflow more precisely by modeling summation as a random walk, specifically a Markov chain with a single absorbing state representing overflow.

To illustrate the idea, consider the summation of integers from the range [-2,2] using an accumulator that can only hold values in [-2,2].
In each step, we randomly select an integer from the range [-2,2] and add it to the accumulator.
We stop when the accumulator overflows out of range [-2,2], i.e., we enter the absorbing state.
Once entered, the process cannot leave the absorbing state.
Hence, the random walk will eventually end as the accumulator is permanently absorbed into an overflow state.
A $6\times6$ transition matrix $P$ represents the probabilities of entering different states given the current sum, with each row summing to 1.
\[
  \Huge{\textit{P = }} \large{\textrm{Input\ State}} 
  \begin{array}{@{}c@{}}
    \rowind{\large{-2}} \\ \rowind{\large{-1}} \\ \rowind{\large{0}} \\ \rowind{\large{1}} \\ \rowind{\large{2}} \\ \rowind{\large{Ovfl}}
  \end{array}
  \mathop{\left[
  \begin{array}{ *{6}{c} }
     \colind{1/5}{\large{-2}}  &  \colind{1/5}{\large{-1}}  &  \colind{1/5}{\large{0}}  & \colind{0}{\large{1}} & \colind{0}{\large{2}} & \colind{2/5}{\large{Ovfl}} \\
    1/5 &  1/5  &  1/5  & 1/5 & 0 & 1/5 \\
     1/5  & 1/5 &  1/5  & 1/5 & 1/5 & 0\\
     0  &  1/5  & 1/5 & 1/5 & 1/5 & 1/5\\
     0  &  0  &  1/5  & 1/5 & 1/5 & 2/5 \\
     0  &  0  &  0  & 0 & 0 & 1
  \end{array}
  \right]}^{
  \begin{array}{@{}c@{}}
    \rowind{\large{Output State}} \\ \mathstrut
  \end{array}
  }
\]

For example, the 5th row represents the probability of different output states given the starting accumulator value of 2.
The value 2 may be summed with either 1 or 2 with probability 2/5 as both are equally likely to be the next state (uniform random draws).
Since 2+1=3 and 2+2=4 both overflow the accumulator, we enter the overflow state (ovfl column) with probability 2/5.
The last row shows that if we start in an overflow state, we will remain in that state surely. 

We can represent the transition matrix \( P \) in a blocked form:
\[ P = \begin{pmatrix}
Q & R \\
0 & I
\end{pmatrix} \]
where $0$ is a zero matrix and $I$ is the identity matrix.
$R$ represents the transitions from transient states to absorbing states, and $Q$ represents the transitions between transient states.
To compute the transition probabilities after $k$ steps, we simply multiply $P$ by itself $k$ times.
\begin{equation*}
    P^k = 
  \begin{pmatrix}
    Q^k & R + QR + ... + Q^k R \\
    0 & I
  \end{pmatrix}
  =
  \begin{pmatrix}
    0 & (I - Q)^{-1}R  \\
    0 & I
  \end{pmatrix}
\end{equation*}
$Q^k = 0$ reflects the fact that the random walk will eventually end, i.e., eventually there is zero probability of being in a non-absorbed state. 
The fundamental matrix of the Markov chain $ N = (I - Q)^{-1}$ represents the expected number of visits to non-absorbing state $j \in [-2,2]$ starting from non-absorbing state $i \in [-2,2]$, before absorption.
The accumulator starts with the value 0, varies across different non-absorbing states with each partial sum, and eventually overflows.
The expected number of steps to reach overflow is simply the sum of the entries in row 3 of $N$, corresponding to the state 0. 
This sum represents the total expected number of visits to all non-absorbing states before absorption, i.e., the total expected number of sums we may perform before overflow.

\begin{figure}[ht]
    \subfloat{\label{fig:contour1}{\includegraphics[width=0.33\linewidth]{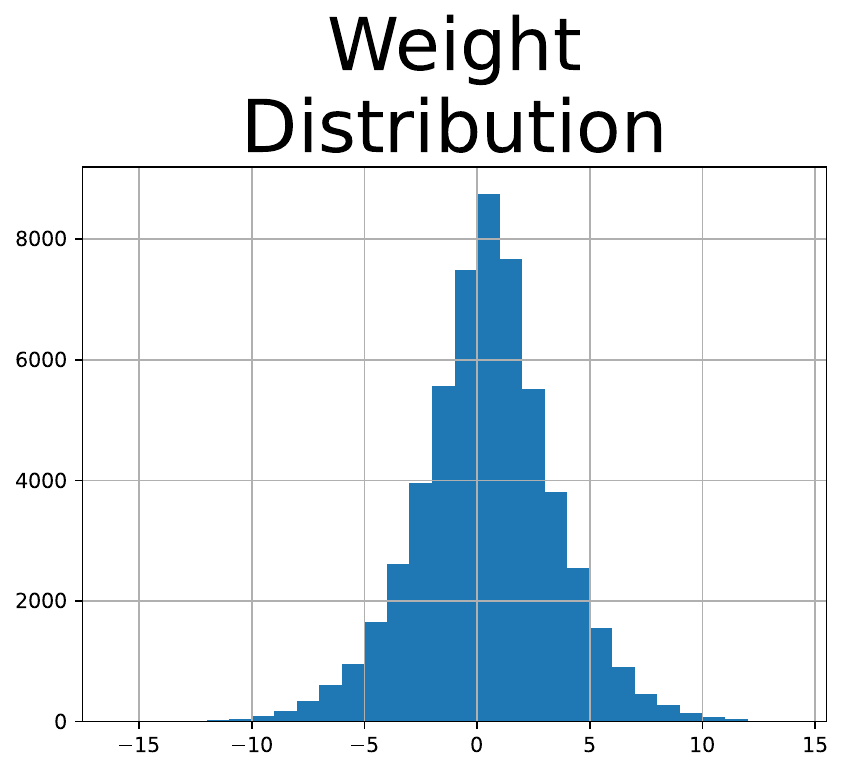}}}
    \subfloat{\label{fig:contour2}{\includegraphics[width=0.34\linewidth]{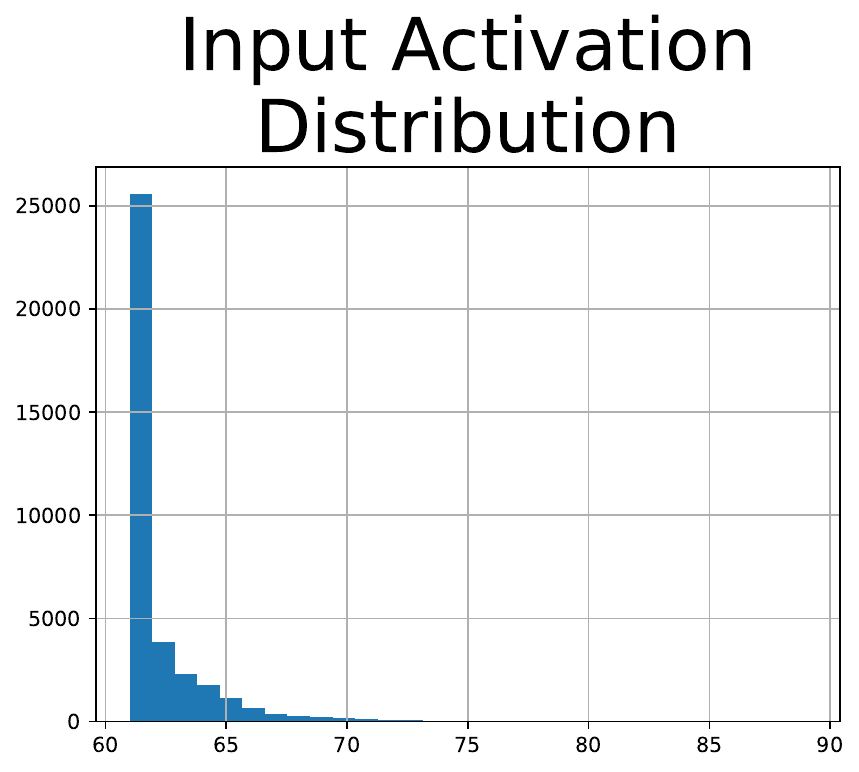}}}
    \subfloat{\label{fig:contour3}{\includegraphics[width=0.33\linewidth]{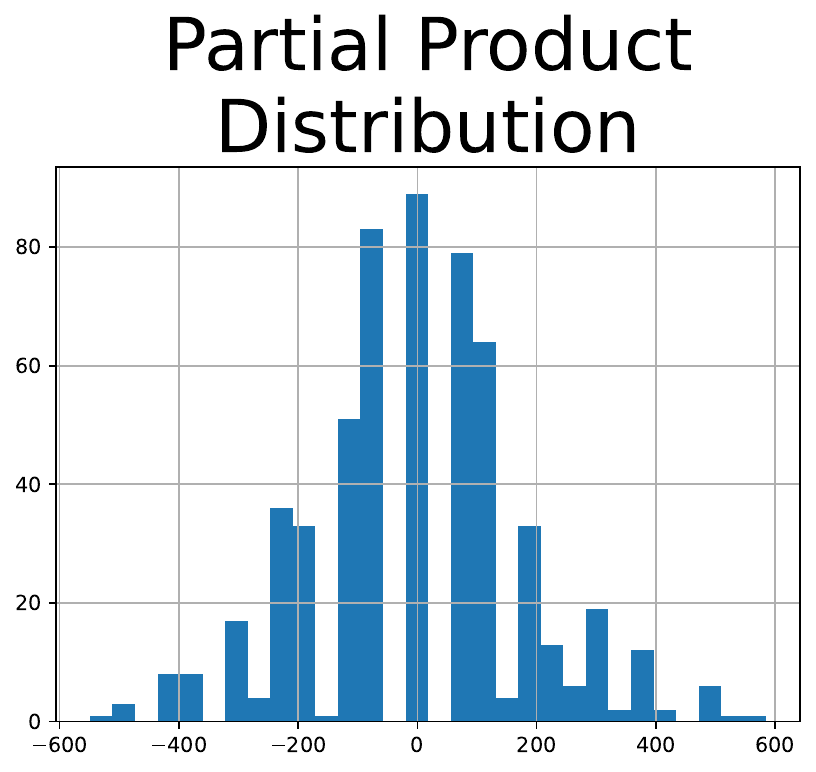}}}\\
    \subfloat{\label{fig:contour4}{\includegraphics[width=\linewidth]{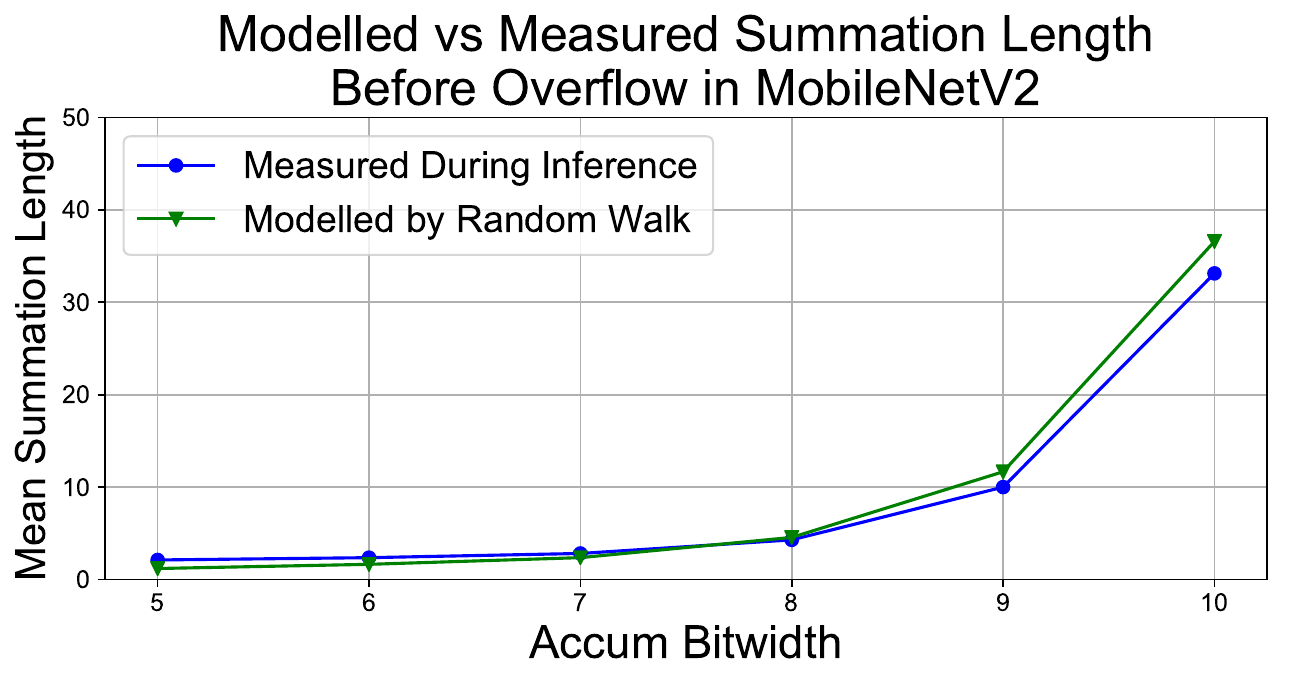}}}
    \vspace{-12pt}
\caption{
Plotting the empirical measured average dot product length versus expected dot product length based on our random walk model.
5-bit Weights follow a normal distribution in the range [-15, 15], while 7-bit activations have a half-normal distribution in the range [0,127] after ReLU. 
Note that the plot shows that with the accumulation bitwidth equal to 10, we do not expect overflow at a summation length of about 32.
In contrast, a naive analysis would conclude that 17 = 5+7+5 bits are required to avoid overflows, noting that $5 = \log_2{32}$.}
\label{fig:model}
\end{figure}

We apply our random walk model to dot product accumulation when executing quantized MobileNetV2 inference on Imagenet1K with 5-bit weights and 7-bit activations \cite{mobilenetv2, imagenet}.
Since weights and activations may deviate slightly from normal, we compute transition probabilities using their empirical distributions during DNN inference.
Figure \ref{fig:model} plots the empirical versus modeled average summation length before overflow in a 1x1 convolution layer in the 13th residual block.
When summing partial products derived from multiplying 5-bit weight and 7-bit activations, we expect that 5+7=12 bit accumulation is required. 
However, Figure \ref{fig:model} shows one may use a narrow 9-bit accumulator to sum 10 values before needing to use a wider accumulator, on average.

\section{Dual-Accumulator MAC Design}
\label{sec:dmac}

In this section, we describe the hardware for dual-multiply-accumulate (dMAC) units, leveraging our observation that the majority of dot product sums do not overflow when using narrow accumulators.
We first introduce the dMAC for integer dot products and then show how this design enables narrow accumulation in FP8.

\subsection{Integer dMAC}

The integer dMAC unit uses a narrow adder (green in Figure 6) for most summations and a wide adder (red) to handle partial sums that overflow the narrow adder.
It has a slightly higher area overhead than a conventional MAC unit, containing two adders and additional overflow handling logic.
However, dMAC consumes significantly less dynamic power by exploiting the low overflow rate in DNN dot products.
In addition, we clock-gate the wider accumulator to reduce dynamic power usage further when not performing wide accumulations.

\begin{figure}[h!]
\centering
\includegraphics[width=\linewidth]{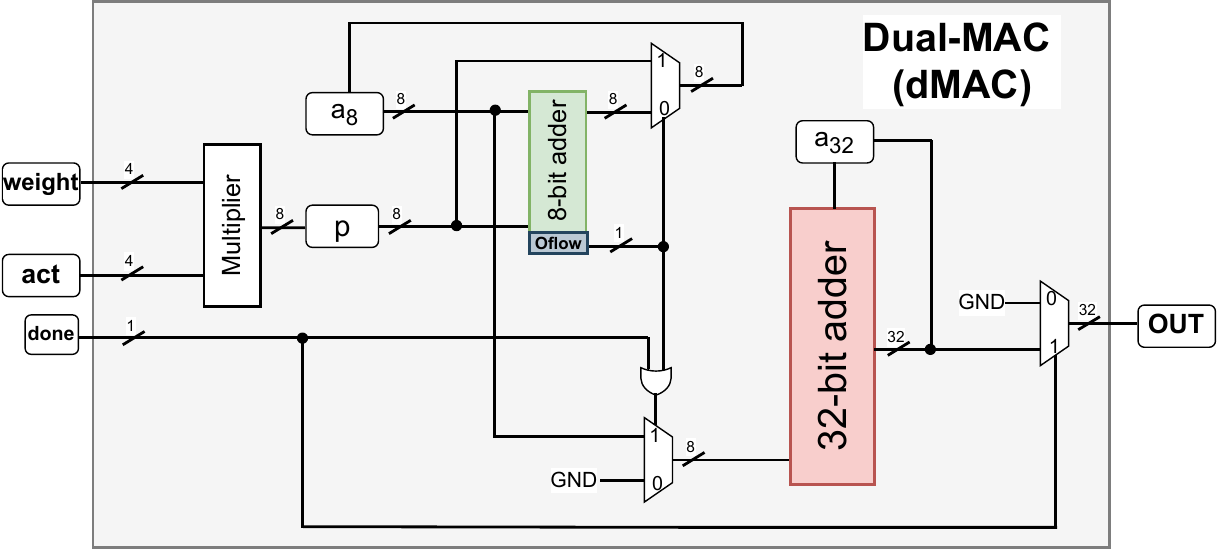}

\caption{
Dual accumulator MAC hardware unit (dMAC) with output-stationary behavior.
In this example, 4-bit weights and data arrive for multiplication and summation with an 8-bit accumulator $a_8$.
If an overflow occurs (oflow = 1), $a_8$ is summed into the wider 32-bit accumulator $a_{32}$, and the 8-bit partial product is written to $a_8$.
Upon completing the dot product (done = 1), we return the sum of $a_8$ and $a_{32}$.
}
\label{fig:dmac}
\end{figure}

Figure \ref{fig:dmac} displays our integer dMAC design when multiplying 4-bit weights and activations using 8-bit and 32-bit adders.
After multiplication, the product $p$ is accumulated in an 8-bit register $a_8$. 
If the 8-bit adder's carry-out overflow flag is set, we accumulate $a_8$ in the wider 32-bit register $a_{32}$ instead and write $p$ to $a_8$. 
Once all the partial products have been accumulated, we add $a_8$ and $a_{32}$ and return the output.

\begin{figure}[h!]
\centering
\includegraphics[width=0.75\linewidth]{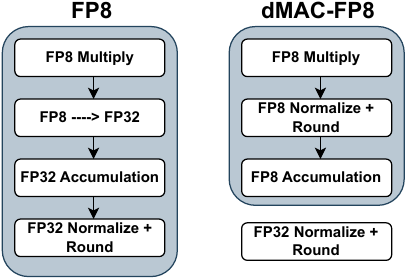}

\caption{
High-level view of operations in our dMAC-FP8 unit versus conventional FP8 MACs. The gray boxes represent the operations that must occur every time a pair of values arrives.
Conventional FP8 incurs overhead from data conversion and wide accumulation and normalization operations.
In contrast, dMAC-FP8 performs the majority of computation in narrower precision while amortizing the cost of normalization across multiple partial summations.
}
\label{fig:fp32}
\end{figure}

\subsection{8-bit Floating-Point dMAC}
\label{sec:fp8}

Existing hardware for FP8 MAC operations accumulate partial products in higher precision such as FP32 \cite{h100, gaudi}
This not only requires the use of wide mantissa adders but also FP8->FP32 data conversions and FP32 normalizations.
Figure \ref{fig:fp32} provides a high-level view of the difference between our hardware and existing FP8 MAC units.
We show that using dMACs for mantissa accumulation can avoid several expensive operations in wide registers while maintaining numerical accuracy.

\begin{figure}[h!]
\centering
\includegraphics[width=0.7\linewidth]{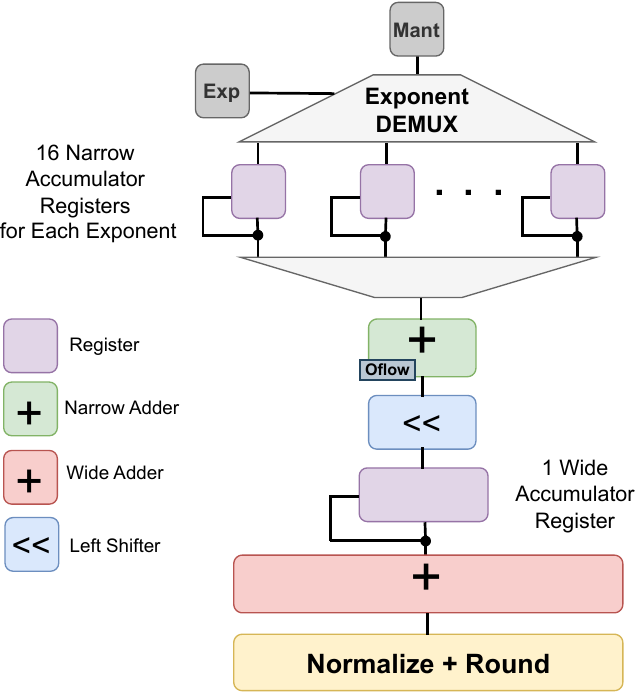}

\caption{
FP8 dMAC hardware unit. We display the FP8 accumulation step in the computation.
A new partial product arrives every cycle. Partial product mantissas are written to one of 16 registers such that they are always accumulated with other values of the same exponent, thereby preventing swamping while enabling the use of the narrow accumulator (green) for most summations.
When the narrow accumulator overflows, it is left-shifted and summed with the wide accumulator (red) to prevent precision loss.
}
\label{fig:fp8}
\end{figure}

Figure \ref{fig:fp8} displays our FP8 dMAC design.
A new weight and activation in the E4M3 format arrive each cycle.
After multiplication and rounding, the partial product sign bit converts the 4-bit mantissa (with leading 1) to 5-bit signed 2's complement.
Using a narrow 5-bit adder, we then accumulate the mantissa into one of 16 5-bit registers based on its 4-bit exponent, which ranges from 0 to 15. 
By accumulating mantissas of the same exponent in the same register, we avoid the shifting operations required when adding two FP8 values with differing exponents while also avoiding numerical error from swamping (see \ref{fig:swamping}).

When the 5-bit adder overflows, we left-shift the accumulator by its exponent and accumulate into a wide 32-bit accumulator.  
Left-shifting by the exponent forces the 5-bit accumulator value to have exponent zero, allowing partial sums with different exponents to be added into the same wide register without error.
Since overflows are rare, dMAC amortizes the shifting cost of mantissa alignment over several summations instead of between each pair of elements as in conventional FP32.
Once the dot product is complete, the values in each accumulator register are left shifted by their respective exponent and summed into the 32-bit accumulator.
This $16\times$ shift+add operation is only performed once per dot product.
Finally, the result is normalized, rounded, and returned.


\subsection{Subnormal Gating}
Zero-gating hardware, such as those proposed in \cite{ye2019zerogating} and \cite{chen2016eyeriss} do not perform multiplication when a zero operand is loaded.
With zero-gating approaches, processing elements are forced to remain idle, saving computation energy.
Meanwhile, zero-skipping approaches such as \cite{han2016eie} and \cite{chen2019eyeriss} avoid loading zero operands from memory altogether.

We exploit the range of possible FP8 products to avoid MAC operations on zeros and small inputs to the dMAC unit.
Given the 256 possible values in the FP8 range, the number of possible partial product pairs in FP8 is ${256 \choose 2} = 32640$.
Of these possible pairs, 1280 lead to product magnitudes that are too small to be represented in the FP8 subnormal range and ultimately round to zero, i.e., $|w\cdot x| < 2^{-9}$.
Moreover, as weights and activations may be normally distributed with mean 0, several partial products will be 0 or close to 0.
We implement logic to check whether input exponents are small enough that the product lies outside the subnormal range.
We then skip these MAC operations for additional dynamic power savings.

\section{Evaluation}
\label{sec:eval}

\begin{figure*}[ht]
    \subfloat{\label{fig:mv2}{\includegraphics[width=0.32\textwidth]{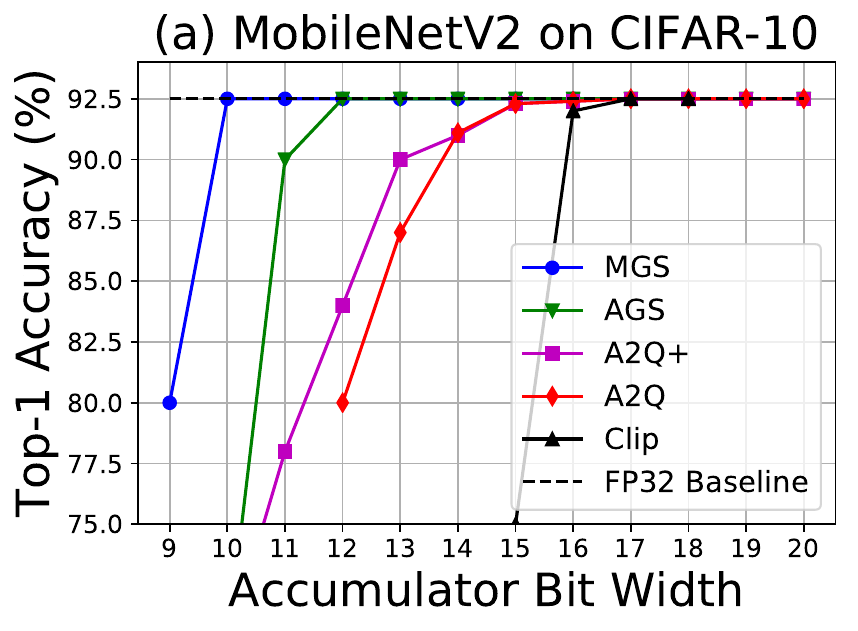}}}\hfill
    \subfloat{\label{fig:resnet}{\includegraphics[width=0.32\textwidth]{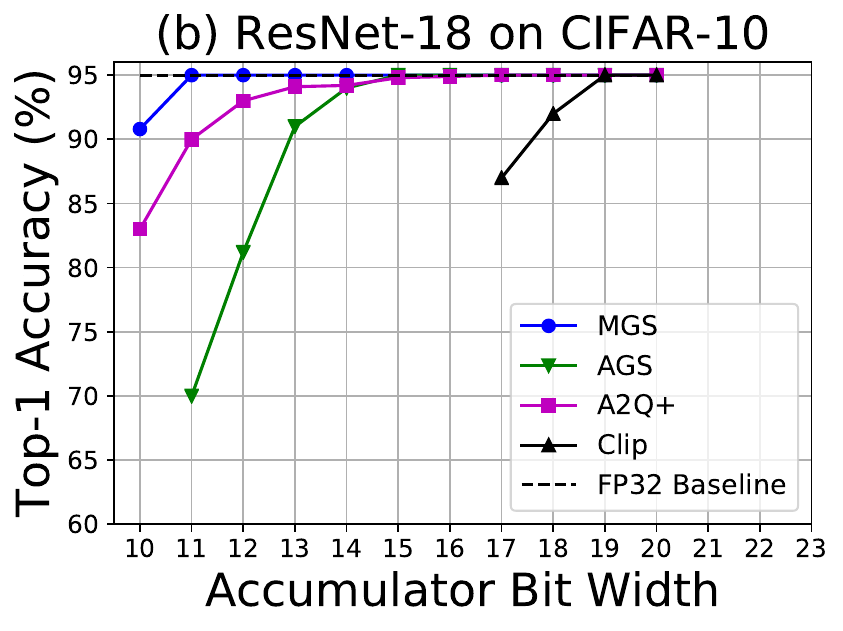}}}\hfill
    \subfloat{\label{fig:vit}{\includegraphics[width=0.32\textwidth]{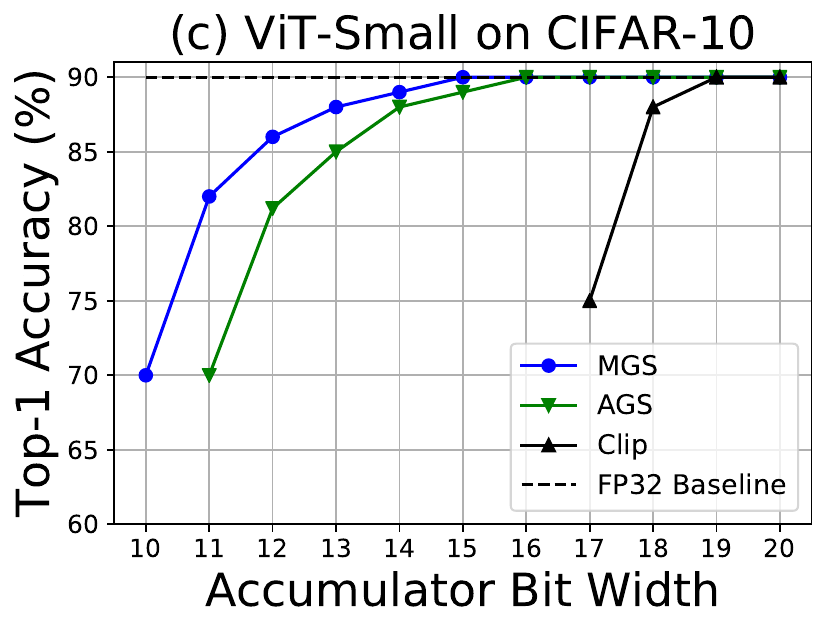}}}\hfill
  
\vspace{-10pt}
\caption{
    Comparing INT8 MGS to SOTA methods for low-precision accumulation during quantized inference using several models.
    We sweep weight and activation bitwidths from 5 to 8 bits while varying the accumulator from 8 to 20 bits.
    We then plot the best-performing models with the lowest required accumulator bitwidth.
    Since MGS uses both narrow and wide accumulators during the dot product, we plot the average accumulator bitwidth when running MGS.
    In principle, MGS can indefinitely reduce the narrow accumulator bitwidth as it always falls back on the wide accumulator.
    However, we stop reducing accumulator bitwidth for MGS when additional reduction increases the average bitwidth (using the wide accumulator more often) and instead start clipping those overflows.
    By using narrow accumulators for the majority of sums, MGS can reduce accumulator bitwidth beyond the SOTA.
} \label{fig:accum}
\vspace{-2mm}
\end{figure*}

Our evaluation is divided into two parts.
In Section \ref{sec:bit}, we compare MGS to state-of-the-art (SOTA) works in accumulator compression when performing inference on various image classification networks.
We show that MGS can significantly reduce accumulator bitwidth while achieving accuracy on par with FP32 baselines, without retraining.
Then, in Section \ref{sec:asic}, we implement dMAC units for INT8 and FP8 in a 7nm node and measure power consumption relative to conventional MACs.
dMACs can reduce inference power consumption by up to 34.1$\%$.

\subsection{FP8 Emulation Library}
Prior works have addressed the difficulty of efficiently profiling overflows due to lack of support in standard deep learning frameworks \cite{a2q, wrapnet}.
We have built a C++/CUDA library to emulate dMAC FP8 on both CPUs and GPUs to run experiments quickly. 
We extend PyTorch's quantization framework with custom linear and convolution layers implementing MGS for INT8 and FP8 quantization to measure the impact on model accuracy.
We unroll dot product computations, allowing users to vary weight, activation, and accumulator bitwidths and evaluate overflow solutions such as MGS, clipping, or wraparound arithmetic.


\subsection{Reducing Accumulator Bitwidth}
\label{sec:bit}

In this section, we evaluate the ability of MGS to enable low-resolution accumulation while maintaining FP32 model accuracy in  MobileNetV2 \cite{mobilenet}, ResNet-18 \cite{resnet}, and ViT \cite{vit} on CIFAR10 \cite{krizhevskycifar10} and ImageNet \cite{imagenet}.
While our evaluation focuses on FP8 inference, MGS may also be applied to other data formats to reduce accumulator bitwidth, e.g., during training with the E5M2 FP8 datatype. 

\subsubsection{8-Bit Integer Quantized DNNs}

We sweep the design space by varying weight and activations from 5 to 8 bits while varying the accumulator bitwidth from 8 to 20.
We select the best-performing models with the lowest required accumulator bitwidth to generate a Pareto frontier.
For models on the frontier, we use our software library to evaluate accuracy compared to SOTA methods A2Q, A2Q+, AGS, and overflow clipping \cite{a2q, a2q+, ags, neon, cmsis1, pulpnn}.

Figure \ref{fig:accum} shows that MGS can push the accumulator bit width lower than A2Q+ while also maintaining task performance.
The magenta lines show that clipping transient overflows within dot products can limit how much we may reduce accumulator bitwidth.
AGS accurately avoids transient overflows but clips persistent overflows, leading to accuracy drops at lower bitwidth where clipping becomes more prevalent.

\subsubsection{8-Bit Floating Point Quantized DNNs}
We evaluate MGS when performing FP8 inference using our target models. 
We employ the E4M3 datatype and fix the narrow accumulator bitwidth at five signed bits. 
Table \ref{tab:acc} shows that MGS accuracy is on par with SOTA methods for FP8.
This is expected since MGS always falls back to using a wide accumulator upon overflow and does not lose accuracy due to swamping.

\begin{center}
\begin{table}
 \caption{Imagenet1K Top-1 Accuracy}
  \label{tab:cpus}
 \centering
  \scalebox{0.9}{
\begin{tabular}{c c c c}
\hline\hline 
Unit & MobileNetV2 & ResNet-18 & ViT-Small \\
\hline 
 Baseline (FP32) & 71.6 & 70.58 & 80.08 \\ 
 INT8 & 69.7  & 68.45  & 79.17  \\ 
 FP8 & 71.04 & 70.12  & 80.02 \\ 
 dMAC & 71.10 & 70.11 & 80.07  \\ 
\hline 
\end{tabular}
}
\label{tab:acc}
\end{table}
\end{center}

\vspace{-12pt}
\subsection{dMAC FPGA Prototyping}
To estimate resource utilization for the MAC units, we prototype the designs on an AMD Virtex-7  VC707 FPGA. Table~\ref{tab:fpga} compares the resource utilization of the FPGA MAC unit implementations in terms of look-up tables (LUTs) and flip-flops (FFs). 
The area of our INT8 dMAC is slightly higher than the conventional INT8 MAC since our unit contains an extra adder and overflow handling logic to use both accumulators.
Although some dMAC designs may have higher resource utilization, they always achieve power savings as detailed in the accurate power, performance, and area characterization of ASIC implementations of the MAC units detailed in Section~\ref{sec:asic}.

\begin{table}[h!]
\caption{FPGA MAC unit resource utilization comparison}
\vspace{-10pt}
 \label{tab:fpga}
 \centering
 \scalebox{1.0}{
\begin{tabular}{|c | c c |}
\hline 
\textbf{Unit} & \begin{tabular}{@{}c@{}}\textbf{FPGA}\\\textbf{LUTs}\end{tabular} & \begin{tabular}{@{}c@{}}\textbf{FPGA}\\\textbf{FFs}\end{tabular}\\
\hline\hline 
 INT8 MAC & 107 & 81\\ 
 \hline
 \textbf{INT8 dMAC} & 126 & 79 \\ 
 \hline
 \hline
 FP8 MAC & 457 & 335 \\ 
  \hline
 \begin{tabular}{@{}c@{}}\textbf{FP8 dMAC (w/o skipping)}\end{tabular}   & 165 & 143 \\
  \hline
 \begin{tabular}{@{}c@{}}\textbf{FP8 dMAC (w/ skipping)}\end{tabular} & 180 & 143 \\
\hline 
\end{tabular}
}
\end{table}

\subsection{dMAC ASIC Physical Implementation}
\label{sec:asic}
\begin{figure}[ht!]
\centering
\includegraphics[width=\linewidth]{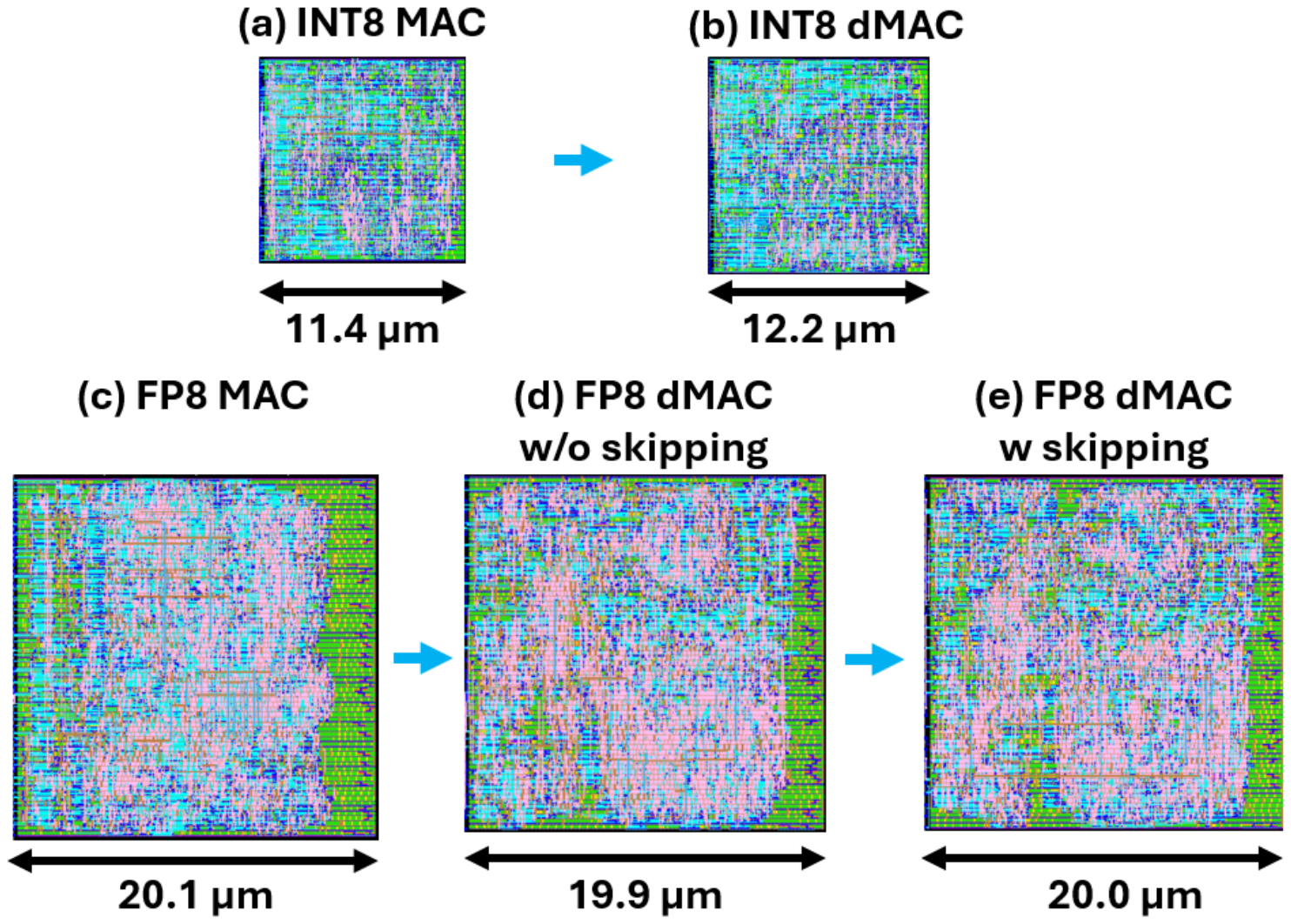}
\vspace{-20pt}
\caption{Layouts and dimensions of ASIC implementations of the MAC units at the 7~nm node: (a) is INT8 MAC, (b) is INT8 dMAC, (c) is FP8 MAC, (d) is FP8 dMAC without skipping, and (e) is FP8 dMAC with skipping.
}
\vspace{-12pt}
\label{fig:layout}
\end{figure}

In addition to comparing the FPGA implementation of the designs, we compare the ASIC implementations for accurate power, performance and area characterization. To compare the performance of each MAC unit, we perform the full physical implementation of the designs at the 7~nm node using the ASAP7 PDK~\cite{clark2016asap7} with a 0.7~V supply voltage. Figure~\ref{fig:layout} shows the layouts and dimensions of the MAC units and their respective areas. To balance switching speed and power consumption, we implement the design using standard threshold voltage transistors targeting a clock frequency of 500~MHz. We synthesize the designs using Cadence Genus~\cite{cadence}, perform implementation using Cadence Innovus, run gate-level simulations with representative weights for each workload using Synopsys VCS~\cite{vcs}, and characterize power consumption based on transient behavior using Cadence Voltus. 

The power comparison of the designs is shown in Table~\ref{tab:asic_power}. As expected, the static leakage power scales with area. Comparing INT8 dMAC with INT8 MAC (with traces from MobileNetV2), we see that although the area for INT8 dMAC is 14.5$\%$ larger and leakage power is 16.4$\%$ higher from additional logic, we still achieve a 15.4$\%$ total power savings. The FP8 dMAC unit without skipping consumes the least area, the FP8 dMAC unit with skipping consumes slightly more area for skipping logic, and the baseline FP8 MAC unit consumes the most area. When comparing total power for FP8 (with traces from ViT), the dMAC achieves 33.6$\%$ and 34.1$\%$ total power savings without and with skipping, respectively. Our zero-skipping approach implements logic to check input exponents before skipping state machine stages within the full MAC operation. Further savings could be made with more aggressive power-saving techniques such as power-gating (where unused parts of the circuit are turned off) ~\cite{shin2010power} or dynamic voltage and frequency scaling~\cite{dvfs}.

\begin{table}[h!]
\caption{ASIC MAC unit power results at 500~MHz showing that 15.4$\%$, 33.6$\%$, and 34.1$\%$ savings are achieved for INT8, FP8 without skipping, and FP8 with skipping, respectively.}
 \label{tab:asic_power}
 \centering
 \scalebox{1.0}{
\begin{tabular}{|c | c c c | c|}
\hline 
\textbf{Unit} & \begin{tabular}{@{}c@{}}\textbf{Dynamic}\\\textbf{Power}\\\textbf{($\mu$W)}\end{tabular} & \begin{tabular}{@{}c@{}}\textbf{Static}\\\textbf{Power}\\\textbf{($\mu$W)}\end{tabular} & \begin{tabular}{@{}c@{}}\textbf{Total}\\\textbf{Power}\\\textbf{($\mu$W)}\end{tabular} & \begin{tabular}{@{}c@{}}\textbf{Power}\\\textbf{Saving}\end{tabular}\\
\hline\hline 
 INT8 MAC & 27.41 & 0.073 & 27.48 & baseline\\ 
 \hline
 INT8 dMAC & 23.16 & 0.085 & 23.25 & $\mathbf{15.4\%}$\\ 
 \hline
 \hline
 FP8 MAC & 97.12 & 0.249 & 97.37 & baseline\\ 
  \hline
 \begin{tabular}{@{}c@{}}FP8 dMAC\\(w/o skipping)\end{tabular}  & 64.44 & 0.226 & 64.66 & $\mathbf{33.6\%}$\\
  \hline
 \begin{tabular}{@{}c@{}}FP8 dMAC\\(w skipping)\end{tabular} & 63.92 & 0.232 & 64.15 & $\mathbf{34.1\%}$\\
\hline 
\end{tabular}
}
\end{table}

\section{Conclusion}
The paper introduced MGS to reduce the required accumulation bitwidth in
performing dot products that form the bulk of DNN computations. Based on the
statistical properties of weight and activation distributions, we can sum many
partial products in reduced precision before overflow occurs. Specifically, MGS uses
a narrow accumulator to accumulate as many values as possible while falling back on a
wider accumulator when the rare overflow occurs. We have designed
dual-multiply-accumulate (dMAC) hardware units that use narrow accumulators for most sums, resulting in a narrower average
accumulator bitwidth compared to prior works. Our dMAC units consume
significantly less power than conventional integer and floating-point MACs that use
wide accumulators for all summations while achieving classification accuracy on par
with high-precision floating-point baselines for multiple image classification tasks.

\bibliographystyle{ACM-Reference-Format}
\bibliography{refs}

\end{document}